\documentclass[12pt,preprint]{aastex}

\newcommand{\myemail}{ehumphreys@cfa.harvard.edu}
\newcommand{\water}{H$_{2}$O }
\newcommand{\gal}{NGC~3079 }
\newcommand{\freq}{183~GHz }

\slugcomment{}

\shorttitle{Extragalactic \water Millimeter Maser Emission}
\shortauthors{Humphreys et al.}

\begin{document}

\title{First Detection of Millimeter/Submillimeter Extragalactic \water Maser Emission}


\author{E. M. L. Humphreys}
\email{\myemail}
\author{L. J. Greenhill}
\email{lgreenhill@cfa.harvard.edu}
\author{M. J. Reid}
\email{mreid@cfa.harvard.edu}
\author{H. Beuther}
 \email{hbeuther@cfa.harvard.edu}
\author{J. M. Moran}
 \email{jmoran@cfa.harvard.edu}
\author{M. Gurwell}
 \email{mgurwell@cfa.harvard.edu}
\author{D. J. Wilner}
\email{dwilner@cfa.harvard.edu}
\author{ P. T. Kondratko}
\email{pkondratko@cfa.harvard.edu}
 
\affil{Harvard-Smithsonian Center for Astrophysics, 60 Garden Street, Cambridge, MA 02138}



\begin{abstract}
We report the first detection of an extragalactic millimeter wavelength 
\water maser  at  \freq towards \gal using the Submillimeter 
Array (SMA), and a tentative submillimeter wave detection of the  439~GHz 
maser towards the same source using the James Clerk Maxwell Telescope (JCMT). 
These \water transitions are known to exhibit maser emission in 
star-forming regions and evolved stars. \gal is a well-studied 
nuclear \water maser source at 22~GHz with a time-variable peak flux 
density in the range 3 -- 12 Jy. 
The \freq \water maser emission, with peak flux density $\sim$0.5~Jy 
(7$\sigma$ detection), also originates from the nuclear region of NGC~3079 and
is spatially coincident with the dust continuum peak at 193 GHz (53 mJy 
integrated).  Peak emission at both 183 and 439~GHz occurs in the same range of velocity
as that covered by the  22~GHz spectrum. 
We estimate the gas to dust ratio of the nucleus of \gal to be $\approx$150, comparable to the 
Galactic value of 160. 
Discovery of maser emission in an active galactic nucleus beyond the long-known 22~GHz transition
opens the possibility of future position-resolved radiative transfer modeling of
accretion disks and outflows $<1$ pc from massive black holes. 
\end{abstract}

\keywords{Masers --- submillimeter ---  galaxies: active --- galaxies: 
nuclei --- galaxies: individual (NGC 3079) ---  techniques: high angular resolution}

\section{Introduction}
\label{s:intro}

Active galactic nuclei (AGN) that exhibit \water maser emission at 22~GHz 
have been objects of particular interest since discovery that the 
maser emission can trace circumnuclear accretion disks, as in NGC~4258 
\citep[]{Miyoshi1995,Greenhill1995a,Greenhill1995b}. 
The presence of maser emission enables direct mapping of AGN 
dynamics within 1 pc of supermassive black holes via Very Long Baseline 
Interferometry (VLBI) \citep[see reviews of][]{Greenhill2002,Maloney2002}.
The additional discovery that \water maser emission can be associated with
shocked gas in jets and outflows has also provided a tracer for a second component 
in AGN \citep[][]{Gallimore1996,Claussen1998,Peck2003,Greenhill2003}.
  
Maser emission at 22~GHz is believed to be predominantly 
collisionally-pumped at gas kinetic temperatures of T$_{k}$ =  400 -- 1000~K, hydrogen 
densities of n(H$_{2}$) =  10$^{8}$ -- 10$^{10}$ cm$^{-3}$, and dust temperatures 
T$_{d}$ $<$ 100 K \citep[e.g.,][]{Deguchi1977,Elitzur1989,Yates1997}. 
In AGN accretion disks, gas heating is thought to arise from X-ray irradiation 
of the disk surfaces,  whereas masers in jets and outflows may be heated via 
shock compressions \citep{Neufeld1994,Kartje1999}.

Since the physical conditions that give rise to emission at 22~GHz also pump other \water 
maser transitions, 
notably those at 183, 321 and 325~GHz \citep[e.g.,][]{Deguchi1977,Neufeld1991,Yates1997},  
22~GHz maser emission is unlikely to be alone in AGN disks, jets and outflows. 
We note that, in galactic regions of star formation and evolved stars, inversion of the 
22 GHz transition is accompanied by inversion of other rotational transitions from the ground
and vibrationally excited \water states  \citep[e.g.,][]{Phillips1980,Waters1980,Menten1989,Menten1990a,
Menten1990b,Feldman1991,Menten1991,Melnick1993}.
 In AGN, some of the masers should occur in broadly the same regions as at 22~GHz, but others could 
 probe uncharted regions of the central engines of AGN, including regions at smaller radii. 
Line ratios of two or more maser transitions originating from the same gas would constrain radiative transfer models 
far better than is now possible.
Previous searches for extragalactic millimeter/submillimeter \water masers have been performed, but 
have been hampered by r.m.s. sensitivities worse than 1~Jy (Menten, private communication).

\gal is an almost edge-on spiral Seyfert 2/LINER galaxy 
with optical heliocentric velocity  of 1125 $\pm$ 6~km~s$^{-1}$ \citep[e.g.,][]{Heckman1980,Ho1997}, 
at a distance of 16~Mpc \citep{Sofue1999}. 
It hosts 
22~GHz \water maser emission at velocities between $\sim$940 and 1350 
km~s$^{-1}$, which are believed to trace a disk-like structure at radii
of 0.4 -- 1.3 pc from a $2 \times 10^6$ $M_{\odot}$ central engine. 
The pc-scale maser disk is aligned with  a dense molecular kpc-scale disk 
\citep{Koda2002} and is located at the apex of a kpc-scale superbubble \citep{Kondratko2005,Cecil2002}.
The bubble is believed to be inflated by a wide-angle outflow \citep{Cecil2001} that is also responsible for the radio continuum lobes along 
the minor axis of the galaxy \citep[e.g.,][]{Duric1988,Trotter1998,Sawada2000,Yamauchi2004}.       
Within a parsec of the nucleus, the relativistic jet appears aligned with the inner wall
of the bubble and may be responsible for weak 22~GHz maser emission away from
the disk \citep[][and references therein]{Kondratko2005}. 
All of these emission sources lie within the primary beam of the SMA observations reported here.


\section{Observations and Data Reduction}

\subsection{SMA Observations at 183~GHz}

We observed the ground-state 
3$_{13}$ $\rightarrow$ 2$_{20}$ transition of para-\water (E$_{u}$/$k$ $\sim$200~K) 
at a rest frequency of 183.310~GHz ($\lambda$: 1.6~mm) 
towards \gal using the SMA\footnote{The SMA is a 
joint project between the Smithsonian Astrophysical Observatory and the Academia Sinica Institute 
of Astronomy and Astrophysics, and is funded by the Smithsonian Institution and the Academia Sinica.}
on 2005 March 1.  
The compact configuration included six antennas, resulting in projected baselines of 8 -- 72 m. 
The SMA receivers operate in double-sideband mode with two 2 GHz sidebands that we tuned to
183.310 and 193.310 GHz (band-center). The correlator provided
continuous coverage of each sideband (in 24 separate chunks of 256 channels) and 0.4 MHz
channel spacing.
The phase center of the observations was  
$\alpha_{2000} =10^h01^m58^s.53$, 
$\delta_{2000} =55^o40'50\rlap{.}''1$ \citep{Cotton1999}. The zenith opacity measured with the NRAO tipping radiometer 
located at the Caltech Submillimeter Observatory was $\tau$(225~GHz)=0.03 - 0.05 throughout.

We calibrated the data within 
the IDL superset MIR 
developed for the Owens Valley Radio Observatory \citep{Scoville1993} and adapted for the SMA; the imaging was performed in MIRIAD.
Data reduction took into account the effect of the pressure-broadened terrestrial \water 
absorption line centered at 183.310~GHz.  
Fortunately, the redshift of the galaxy reduces the line frequency of the maser
transition by $\sim$0.7~GHz,  greatly lowering the corresponding 
atmospheric opacity.
For bandpass calibration we used Jupiter and Saturn.
Phase calibration was performed via frequent observations 
of the quasar 0923+392 about 18$\rm ^o$ from the phase center. 
We performed the amplitude calibration separately for upper and lower sideband
data. For the upper sideband, we used 0923+392;
for lower sideband data, we performed amplitude calibration on each of the 24 spectral
chunks separately using Saturn and Jupiter. The calibration of the lower sideband both as a function
of elevation and frequency was designed to reduce the effect of variation
in terrestrial \water absorption
across the sideband. We derived the flux scale using observations of Callisto with an  
estimated accuracy of 20\%. 
Single sideband system temperatures corrected to the top of the atmosphere
were between 400 and 850 K. 

\subsection{JCMT Observations at 439~GHz}

We observed the ground-state 
6$_{43}$ $\rightarrow$ 5$_{50}$ transition of ortho-\water (E$_{u}$/$k$ $\sim$1100 K) 
at a rest frequency of 439.151~GHz ($\lambda$:~0.7~mm)  towards \gal using the 15 meter JCMT
on 2004 November 18  
in $\tau_{225}$=0.041 -- 0.045 weather. 
The instantaneous bandwidth was 1840~MHz, corresponding to $\sim$1200~km~s$^{-1}$, 
with a channel spacing of 1250 kHz or 0.85~km~s$^{-1}$.
The integration time on source was 2~hrs in beam-switching mode with a throw of  60$''$. The average T$_{sys}$ during the observations
was 2650~K. We reduced the data using the Starlink spectral-line reduction software, {\sc SpecX}.

\section{Results}
\label{s:results}

\begin{figure}
\includegraphics[angle=-90,scale=.7]{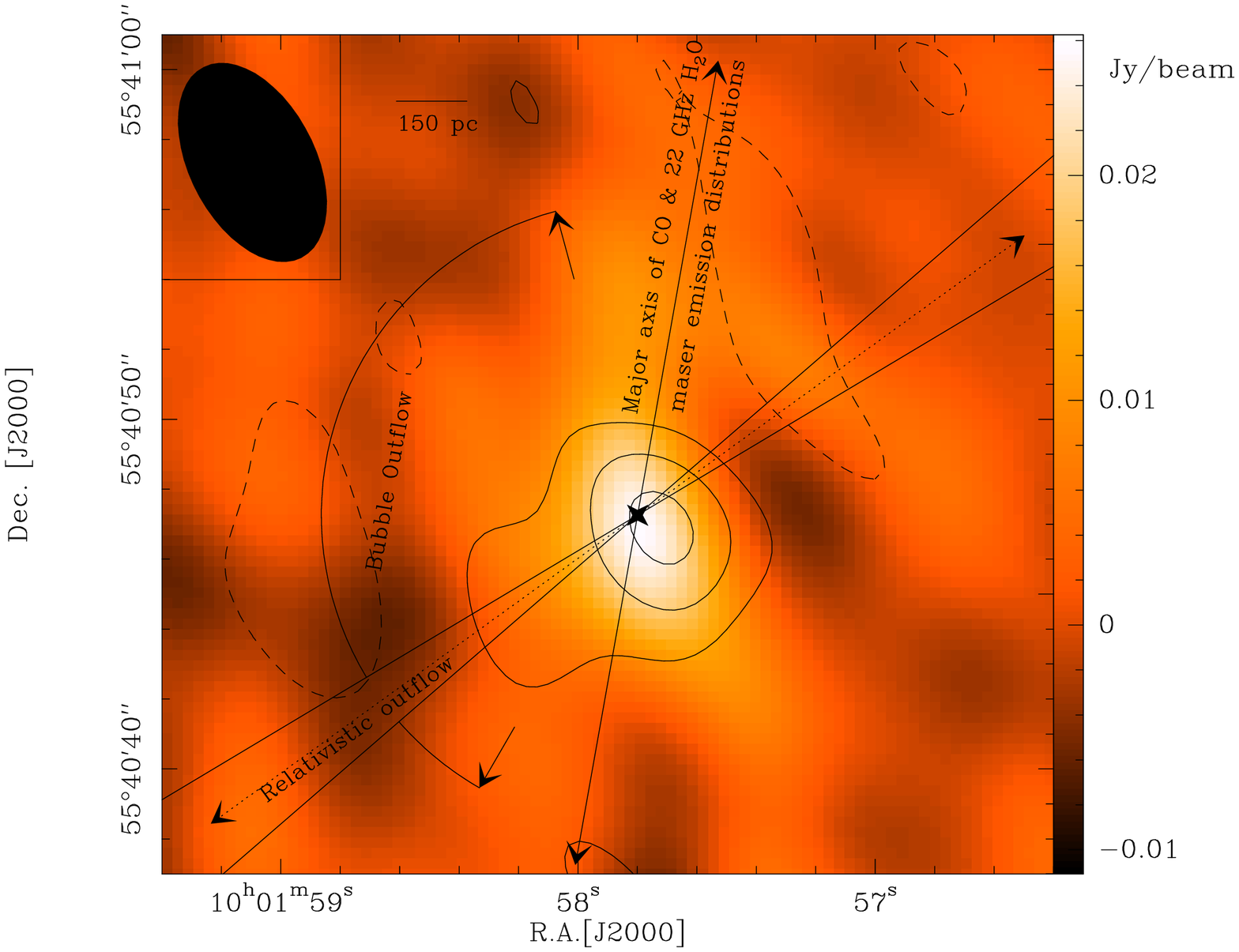}
\caption{\label{f:continuum} 
{\it (filled color contours)} Millimeter continuum image of \gal at 193~GHz observed 
with the SMA. Peak flux density is 27~mJy~bm$^{-1}$ with a  1$\sigma$ r.m.s.
of 2.5~mJy~bm$^{-1}$; integrated continuum flux density is 53~mJy. 
{\it (line contours)} 183~GHz maser emission at 1017 km~s$^{-1}$ (optical, heliocentric).
 Peak maser flux density in the image is 0.39~Jy~bm$^{-1}$. Positive emission is shown 
in solid lines and negative features in dashed contours. 
Contours are $k \times 2\sigma$, where $k=-1,1,2,3$ and 1$\sigma$=56~mJy~bm$^{-1}$.  
The black cross marks the position of peak 22~GHz \water maser emission at 956~km~s$^{-1}$: 
$\alpha_{2000} =10^h01^m57^s.802\pm0.001$, $\delta_{2000} =55^o40'47\rlap{.}''26\pm0.01$ \citep{Kondratko2005}. 
The black solid arrow indicates the P.A.
($\sim$-10$\rm ^o$ east of north) of the pc-scale  22~GHz \water maser disk \citep{Kondratko2005} and of 
the kpc-scale CO disk  \citep{Koda2002}. The
dashed arrow marks the direction of relativistic outflow with uncertainty depicted by bracketing solid
lines of P.A. 126$\rm ^o\pm5^o$ 
\citep{Trotter1998,Sawada2000,Middelberg2004,Kondratko2005}.
The arc to the east denotes the opening angle of the kpc-scale superbubble outflowing from the galactic
nucleus \citep[e.g.,][]{Cecil2002}.  The SMA synthesized beam has
FWHM of 5$\rlap{.}''$9 x 3$\rlap{.}''$4 (upper sideband) and 6$\rlap{.}''$1 x 3$\rlap{.}''$6 (lower sideband, shown top left), with a P.A. of 27$\rm ^o$ in both cases. 
}
\end{figure}

We detected millimeter (193~GHz) continuum emission with a peak signal to noise ratio (S/N) of 
$\sim$10 (Fig.~\ref{f:continuum}) at $\alpha_{2000} =10^h 01^m 57^s .80\pm0.02$, 
$\delta_{2000} = 55 \rm ^o 40' 46\rlap{.}''9\pm0.3$ (statistical uncertainties). 
Systematic uncertainties of ($0\rlap{.}^{s}02$,$0\rlap{.}''3$) were estimated through measurement of
an apparent position for 1150+497 against the phase calibrator 0923+392. These place the 
peak of the millimeter continuum within 1$\sigma$ of the
22 GHz maser position.  
We partially resolved the nucleus in the north-south direction and the elongation is approximately 
parallel to the position angles of both the pc-scale accretion disk traced by
22 GHz maser emission and the kpc-scale molecular disk detected using  e.g., CO \citep{Koda2002}. 

We detected 183~GHz \water emission with a peak S/N of $\sim$7 in both the 
real part of the 
amplitude spectrum and the interferometer phase spectrum (Figs.~\ref{f:continuum} and~\ref{f:line}). 
The peak flux density is 0.55 Jy at 1017 km~s$^{-1}$.
The line covers 9 frequency channels and overlaps a persistent Doppler component appearing in published 22 GHz maser spectra for dates 1998 May 8, 2000 March 18 where it is especially prominent at $\sim$2.5 Jy \citep[both][]{Hagiwara2002},  2001 March 23 \citep{Kondratko2005} and
 2002 April 12 \citep{Braatz2003}. We note possible detection of additional spectral features at 1208 (5 channels wide) 
and 1265 km~s$^{-1}$ (4 channels wide). Both features appear in spectra of real amplitude and phase (near zero degrees, which indicates positions similar to the 1017 km~s$^{-1}$  emission), and both lie within the velocity interval of 22 GHz emission. 
Note that the correspondence of features between 22 and 183 GHz
observations is not precise and is also not expected to be true for
all features under certain pumping scenarios.
The peak maser emission is at $\alpha_{2000} =10^h01^m57^s.75\pm 0.03$, $\delta_{2000}=55^o40'46\rlap{.}''7\pm0.4$, which is within 1$\sigma$ of the
193 GHz continuum peak. At our resolution, we cannot determine
whether 183~GHz emission originates from the
disk or outflow, or from both. We note that in addition to
tracing a disk, a subset of 22 GHz
maser emission is associated with the outflow within a few parsecs of
the nucleus \citep{Kondratko2005}. 22~GHz emission at around
 1000 and 1200 km~s$^{-1}$ is present both in the disk and outflow. 
Possible angular extension of the \freq   emission to the southeast 
is marginally detected (2$\sigma$), which lies along the axis of the relativistic flow \citep[e.g.,][]{Kondratko2005}. We note that \citet{Hagiwara2004} detect OH absorption associated
with the outflow at an overlapping velocity, and that we cannot rule out
shocks in circumnuclear material as an origin for 183~GHz emission.

We obtained a tentative detection (5$\sigma$) of the 439~GHz maser 
at 1157 km~s$^{-1}$ (Fig.~\ref{f:439ghz}). With peak S$_{\nu}$ $\sim$2.5 Jy, 
439~GHz emission is of comparable strength to 
typical values for peak 22~GHz emission towards NGC~3079 and occurs
within the 22~GHz velocity interval. 


\section{Discussion}
\label{s:discussion} 

\subsection{Millimeter Dust Emission}  

The millimeter continuum of NGC~3079 is probably dominated by thermal emission from optically thin dust. 
As in the 1.2 mm observations of \citet{Braine1997} at 11$''$ resolution, we find dust concentrated in the nuclear region of NGC~3079 (Fig.~\ref{f:continuum}) and place an upper limit of 8~mJy (3$\sigma$) on emission associated with the radio lobes $\sim$1 kpc from the emission peak. 
\citet{Braine1997} estimates the dust temperature to be $\sim$30 K at kpc radii. However, in two other AGN the presence of hotter dust at 100 times smaller radii, of $\sim$300 K, has been inferred from VLT observations 
\citep{Jaffe2004,Prieto2004}. Adopting a dust temperature of 100 K for \gal on
intermediate scales and a dust opacity index $\beta=2$, we estimate the enclosed
dust mass of the inner 200 pc is $M_{dust}$ $\sim$ 10$^{6.3\pm0.8}$ M$_{\odot}$ using the method described in \citet{Hildebrand1983}. Using CO Nobeyama observations by \citet{Koda2002}, who find a gas mass of  $M_{gas}$ $\sim$10$^{8.5}$ M$_{\odot}$ within the central 150 pc, we estimate a gas-to-dust ratio  of $150$, which is comparable to the Galactic value of 160 and
to the ratios derived in bright local galaxies \citep{Dunne2000}.

\subsection{183~GHz Emission}

The 183~GHz emission we detect in the nucleus of NGC~3079 is likely from a maser process. It arises from one, and possibly more, narrow 
unresolved features in the same range of velocity as that of known 22~GHz maser 
emission. From radiative transfer models we know that the same high densities and 
temperature conditions that produce maser emission at 22~GHz 
also strongly invert the 183~GHz \water transition. 
Noting that
the 22 GHz feature at 1017 km~s$^{-1}$ varies between 0.4 -- 2.6 Jy (Hagiwara 
et al. 2002), and assuming an ortho:para \water ratio of 3:1, pumping
models could reproduce likely line ratios \citep[e.g.,][]{Yates1997}.
Narrow linewidths ($<$8 km~s$^{-1}$) 
at 183~GHz, small compared with the characteristic rotational and apparent random velocities for 
molecular gas in this nucleus \citep{Kondratko2005}, also indicate that emitting regions should be 
compact. 
Our beam size is too large to determine whether emission is thermal 
or maser based on brightness temperature arguments. For the emission to be thermal, with a brightness
temperature of $\leq$ 2000 K such that \water molecules are not dissociated, the emitting region 
must be $>$ 0$\rlap{.}''$1 (10 pc). 

\subsection{439~GHz Emission}

\begin{figure*}
\includegraphics[angle=-90,scale=.7]{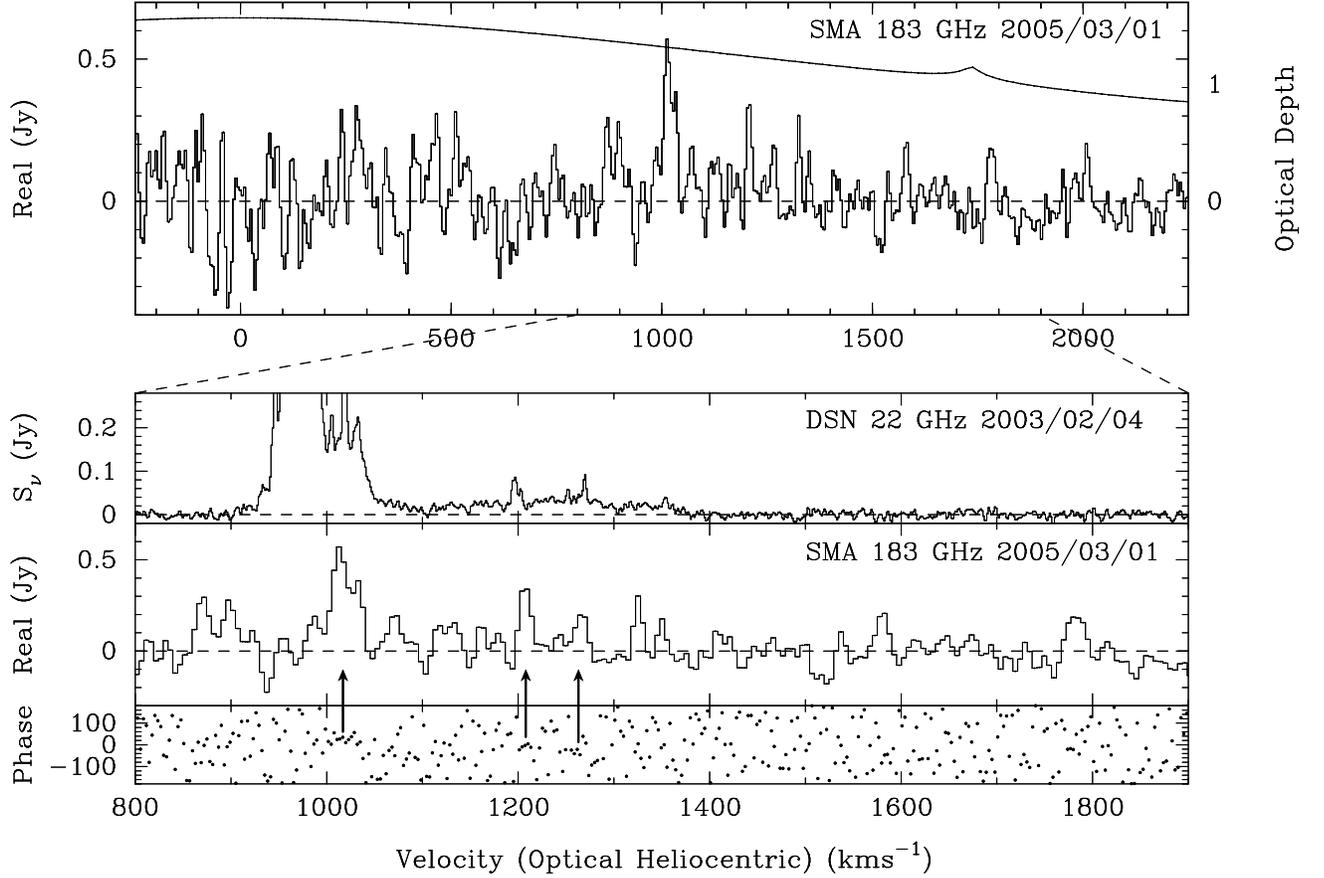}
\caption{\label{f:line} 183~GHz maser spectrum towards the nucleus of \gal corrected for 
atmospheric absorption obtained with the SMA. {\it (top panel)} Real part of the
spectrum exhibiting a $\sim$0.55 Jy feature at a velocity of 1017 km~s$^{-1}$. The 
channel spacing is 5 km~s$^{-1}$.  Atmospheric opacity for a zenith angle of 50$\rm ^o$ 
is shown by the dotted line; peaks at 1850 and 2150 km~s$^{-1}$ are due to ozone. Note 
that the noise rises at lower velocities due to the effect of atmospheric absorption. 
{\it (upper bottom panel)} 22 GHz spectrum obtained with a DSN 70 m antenna (Kondratko, private
communication)  between 800 -- 1900  km~s$^{-1}$. {\it (middle bottom)}
Real part of the 183 GHz SMA
spectrum between 800 -- 1900  km~s$^{-1}$.  
{\it (lower bottom)} Corresponding fringe phase.
Tentative maser features at 1208 and 1265 km~s$^{-1}$ 
are also noted with arrows. 
}
\end{figure*}

\begin{figure}
\includegraphics[angle=-90,scale=.7]{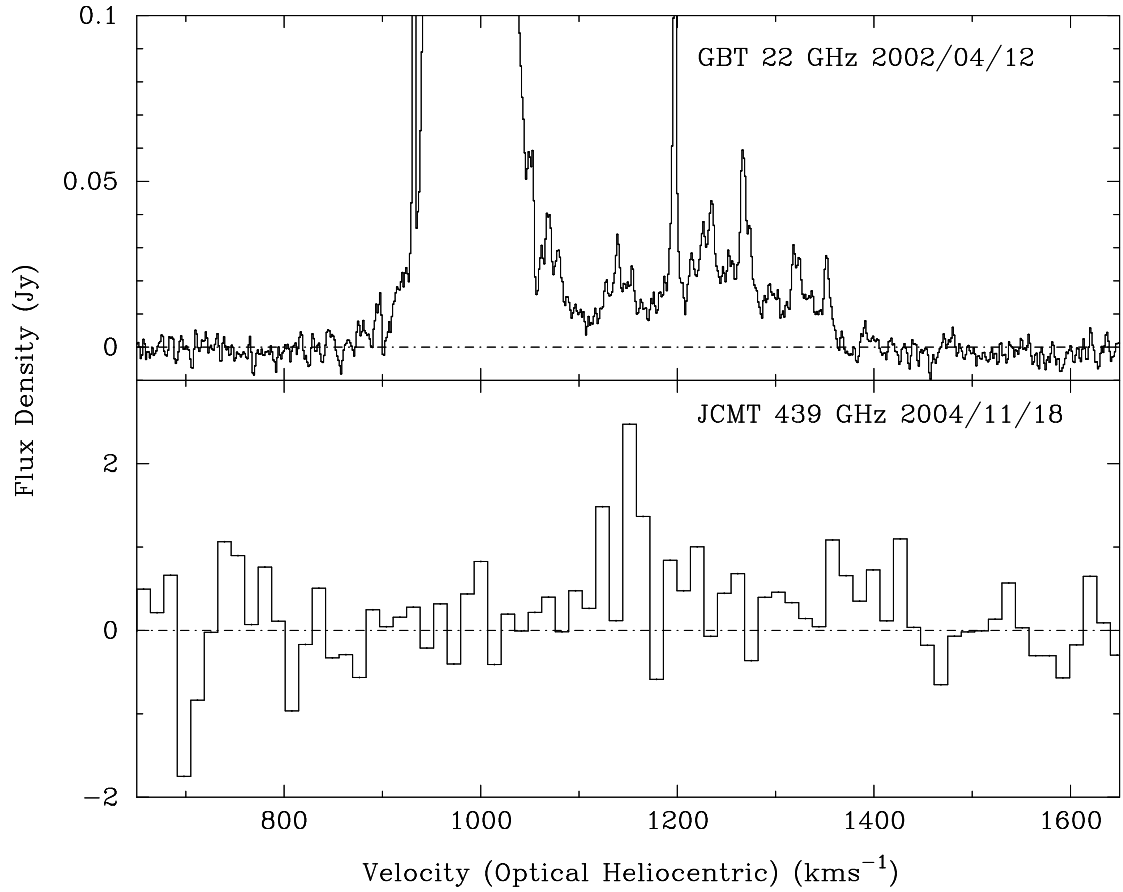}
\caption{\label{f:439ghz} {\it (top)} 22~GHz spectrum obtained using the GBT from 
\citet{Braatz2003}.
{\it (bottom)} 439~GHz \water emission towards \gal observed with the JCMT. 
Peak flux density is 2.5~Jy, and the 1$\sigma$ r.m.s. is 0.5~Jy. 
The channel spacing is 13.7~km~s$^{-1}$. 
}
\end{figure}

439 GHz 
emission is probably also due to maser amplification. 
Emission occurs over a narrow velocity range (Fig.~\ref{f:439ghz}), 
and conditions for strong 439 GHz inversion are likely to exist in 
the galactic nucleus.
The 22~GHz spectrum peaks to the blue of the v$_{sys}$ of 1125 km~s$^{-1}$, yet
the 439~GHz spectrum peaks slightly above v$_{sys}$ at 1157 km~s$^{-1}$ (Fig.~\ref{f:439ghz}). 
We ascribe this difference to the different  pump mechanisms
operating for each maser. Whereas 22~GHz emission is collisionally-pumped 
and is increasingly quenched by local dust temperatures $>$100~K, 
the 439~GHz maser is predominantly radiatively-pumped 
and is most strongly inverted in the presence of $T_{d}=300~K$, $T_{k}=400~K$ 
and n$(H_{2})$ = 5 x 10$^{9}$ cm$^{-3}$ \citep{Yates1997}.  
Dust temperatures of this order 
and higher are reasonable in the central parsecs of the nucleus 
\citep{Jaffe2004,Prieto2004}.
Peak 439~GHz emission (1157 km~s$^{-1}$) is therefore likely to originate from a hot, 
dusty region around 
the central engine, possibly at radii $\ll$1 pc, separated spatially from the gas emitting 
most strongly at 22~GHz (956 km~s$^{-1}$). At 22 GHz, features at these velocities are indeed
separated spatially by $\sim$1 pc projected \citep{Kondratko2005}.
The observations in Fig.~\ref{f:439ghz} were taken on different dates, 
and we note that
the 22~GHz feature near 1157 km~s$^{-1}$ is not present at all epochs (e.g., in
Fig.~\ref{f:line}). 
However, the feature line ratio in Fig.~\ref{f:439ghz} of $\sim$100 
is accommodated by the vicinity of hot dust \citep{Yates1997}. 

\vspace*{-0.55cm}

\acknowledgments
We thank the all the SMA and JCMT staff.  
We thank S. Paine for opacity data,   J. Braatz for 22 GHz data
and G. Melnick for useful discussions. 
H. B. acknowledges support from the Emmy-Noether Program of the
Deutsche Forschungsgemeinschaft (DFG-BE2578/1).

\clearpage

\clearpage

\clearpage

\begin{thebibliography}{50}
\expandafter\ifx\csname natexlab\endcsname\relax\def\natexlab#1{#1}\fi

\bibitem[{{Braatz} {et~al.}(2003){Braatz}, {Wilson}, {Henkel}, {Gough}, \&
  {Sinclair}}]{Braatz2003}
{Braatz}, J.~A., {Wilson}, A.~S., {Henkel}, C., {Gough}, R., \& {Sinclair}, M.
  2003, \apjs, 146, 249

\bibitem[{{Braine} {et~al.}(1997){Braine}, {Guelin}, {Dumke}, {Brouillet},
  {Herpin}, \& {Wielebinski}}]{Braine1997}
{Braine}, J., {Guelin}, M., {Dumke}, M., {Brouillet}, N., {Herpin}, F., \&
  {Wielebinski}, R. 1997, \aap, 326, 963

\bibitem[{{Cecil} {et~al.}(2002){Cecil}, {Bland-Hawthorn}, \&
  {Veilleux}}]{Cecil2002}
{Cecil}, G., {Bland-Hawthorn}, J., \& {Veilleux}, S. 2002, \apj, 576, 745

\bibitem[{{Cecil} {et~al.}(2001){Cecil}, {Bland-Hawthorn}, {Veilleux}, \&
  {Filippenko}}]{Cecil2001}
{Cecil}, G., {Bland-Hawthorn}, J., {Veilleux}, S., \& {Filippenko}, A.~V. 2001,
  \apj, 555, 338


\bibitem[{{Claussen} {et~al.}(1998){Claussen}, {Diamond}, {Braatz}, {Wilson},
  \& {Henkel}}]{Claussen1998}
{Claussen}, M.~J., {Diamond}, P.~J., {Braatz}, J.~A., {Wilson}, A.~S., \&
  {Henkel}, C. 1998, \apjl, 500, L129

\bibitem[{{Cotton} {et~al.}(1999){Cotton}, {Condon}, \&
  {Arbizzani}}]{Cotton1999}
{Cotton}, W.~D., {Condon}, J.~J., \& {Arbizzani}, E. 1999, \apjs, 125, 409


\bibitem[{{Deguchi}(1977)}]{Deguchi1977}
{Deguchi}, S. 1977, \pasj, 29, 669

\bibitem[{{Dunne} {et~al.}(2000){Dunne}, {Eales}, {Edmunds}, {Ivison},
  {Alexander}, \& {Clements}}]{Dunne2000}
{Dunne}, L., {Eales}, S., {Edmunds}, M., {Ivison}, R., {Alexander}, P., \&
  {Clements}, D.~L. 2000, \mnras, 315, 115

\bibitem[{{Duric} \& {Seaquist}(1988)}]{Duric1988}
{Duric}, N. \& {Seaquist}, E.~R. 1988, \apj, 326, 574

\bibitem[{{Elitzur} {et~al.}(1989){Elitzur}, {Hollenbach}, \&
  {McKee}}]{Elitzur1989}
{Elitzur}, M., {Hollenbach}, D.~J., \& {McKee}, C.~F. 1989, \apj, 346, 983

\bibitem[{{Feldman} {et~al.}(1991){Feldman}, {Matthews}, {Cunningham},
  {Hayward}, {Wade}, {Amano}, \& {Scappini}}]{Feldman1991}
{Feldman}, P.~A., {Matthews}, H.~E., {Cunningham}, C.~T., {Hayward}, R.~H.,
  {Wade}, J.~D., {Amano}, T., \& {Scappini}, F. 1991, \jrasc, 85, 191

\bibitem[{{Gallimore} {et~al.}(1996){Gallimore}, {Baum}, {O'Dea}, {Brinks}, \&
  {Pedlar}}]{Gallimore1996}
{Gallimore}, J.~F., {Baum}, S.~A., {O'Dea}, C.~P., {Brinks}, E., \& {Pedlar},
  A. 1996, \apj, 462, 740

\bibitem[{{Greenhill}(2002)}]{Greenhill2002}
{Greenhill}, L. 2002, in IAU Symposium, 381


\bibitem[{{Greenhill} {et~al.}(2003){Greenhill}, {Booth}, {Ellingsen},
  {Herrnstein}, {Jauncey}, {McCulloch}, {Moran}, {Norris}, {Reynolds}, \&
  {Tzioumis}}]{Greenhill2003}
{Greenhill}, L.~J., {et~al.} 2003, \apj, 590, 162

\bibitem[{{Greenhill} {et~al.}(1995{\natexlab{a}}){Greenhill}, {Henkel},
  {Becker}, {Wilson}, \& {Wouterloot}}]{Greenhill1995a}
{Greenhill}, L.~J., {Henkel}, C., {Becker}, R., {Wilson}, T.~L., \&
  {Wouterloot}, J.~G.~A. 1995{\natexlab{a}}, \aap, 304, 21

\bibitem[{{Greenhill} {et~al.}(1995{\natexlab{b}}){Greenhill}, {Jiang},
  {Moran}, {Reid}, {Lo}, \& {Claussen}}]{Greenhill1995b}
{Greenhill}, L.~J., {Jiang}, D.~R., {Moran}, J.~M., {Reid}, M.~J., {Lo}, K.~Y.,
  \& {Claussen}, M.~J. 1995{\natexlab{b}}, \apj, 440, 619

\bibitem[{{Hagiwara} {et~al.}(2002){Hagiwara}, {Henkel}, {Sherwood}, \&
  {Baan}}]{Hagiwara2002}
{Hagiwara}, Y., {Henkel}, C., {Sherwood}, W.~A., \& {Baan}, W.~A. 2002, \aap,
  387, L29

\bibitem[{{Hagiwara} {et~al.}(2004){Hagiwara}, {Kl\"{o}ckner}, \&
  {Baan}}]{Hagiwara2004}
{Hagiwara}, Y., {Kl\"{o}ckner}, H.-R., \& {Baan}, W.~A. 2004, \mnras,
  353, 1055

\bibitem[{{Heckman}(1980)}]{Heckman1980}
{Heckman}, T.~M. 1980, \aap, 87, 152

\bibitem[{{Hildebrand}(1983)}]{Hildebrand1983}
{Hildebrand}, R.~H. 1983, \qjras, 24, 267

\bibitem[{{Ho} {et~al.}(1997){Ho}, {Filippenko}, \& {Sargent}}]{Ho1997}
{Ho}, L.~C., {Filippenko}, A.~V., \& {Sargent}, W.~L.~W. 1997, \apjs, 112, 315


\bibitem[{{Jaffe} {et~al.}(2004){Jaffe}, {Meisenheimer}, {R{\" o}ttgering},
  {Leinert}, {Richichi}, {Chesneau}, {Fraix-Burnet}, {Glazenborg-Kluttig},
  {Granato}, {Graser}, {Heijligers}, {K{\" o}hler}, {Malbet}, {Miley},
  {Paresce}, {Pel}, {Perrin}, {Przygodda}, {Schoeller}, {Sol}, {Waters},
  {Weigelt}, {Woillez}, \& {de Zeeuw}}]{Jaffe2004}
{Jaffe}, W., {et~al.} 2004, \nat, 429, 47

\bibitem[{{Kartje} {et~al.}(1999){Kartje}, {K{\" o}nigl}, \&
  {Elitzur}}]{Kartje1999}
{Kartje}, J.~F., {K{\" o}nigl}, A., \& {Elitzur}, M. 1999, \apj, 513, 180


\bibitem[{{Koda} {et~al.}(2002){Koda}, {Sofue}, {Kohno}, {Nakanishi},
  {Onodera}, {Okumura}, \& {Irwin}}]{Koda2002}
{Koda}, J., {Sofue}, Y., {Kohno}, K., {Nakanishi}, H., {Onodera}, S.,
  {Okumura}, S.~K., \& {Irwin}, J.~A. 2002, \apj, 573, 105


\bibitem[{{Kondratko} {et~al.}(2005){Kondratko}, {Greenhill}, \&
  {Moran}}]{Kondratko2005}
{Kondratko}, P.~T., {Greenhill}, L.~J., \& {Moran}, J.~M. 2005, \apj, 618, 618

\bibitem[{{Maloney}(2002)}]{Maloney2002}
{Maloney}, P.~R. 2002, PASA, 19, 401

\bibitem[{{Melnick} {et~al.}(1993){Melnick}, {Menten}, {Phillips}, \&
  {Hunter}}]{Melnick1993}
{Melnick}, G.~J., {Menten}, K.~M., {Phillips}, T.~G., \& {Hunter}, T. 1993,
  \apjl, 416, L37

\bibitem[{{Menten} \& {Melnick}(1989)}]{Menten1989}
{Menten}, K.~M. \& {Melnick}, G.~J. 1989, \apjl, 341, L91

\bibitem[{{Menten} \& {Melnick}(1991)}]{Menten1991}
---. 1991, \apj, 377, 647

\bibitem[{{Menten} {et~al.}(1990{\natexlab{a}}){Menten}, {Melnick}, \&
  {Phillips}}]{Menten1990a}
{Menten}, K.~M., {Melnick}, G.~J., \& {Phillips}, T.~G. 1990{\natexlab{a}},
  \apjl, 350, L41

\bibitem[{{Menten} {et~al.}(1990{\natexlab{b}}){Menten}, {Melnick}, {Phillips},
  \& {Neufeld}}]{Menten1990b}
{Menten}, K.~M., {Melnick}, G.~J., {Phillips}, T.~G., \& {Neufeld}, D.~A.
  1990{\natexlab{b}}, \apjl, 363, L27

\bibitem[{{Middelberg} {et~al.}(2004){Middelberg}, {Roy}, {Nagar}, {Krichbaum},
  {Norris}, {Wilson}, {Falcke}, {Colbert}, {Witzel}, \&
  {Fricke}}]{Middelberg2004}
{Middelberg}, E., {et~al.} 2004, \aap, 417, 925


\bibitem[{{Miyoshi} {et~al.}(1995){Miyoshi}, {Moran}, {Herrnstein},
  {Greenhill}, {Nakai}, {Diamond}, \& {Inoue}}]{Miyoshi1995}
{Miyoshi}, M., {Moran}, J., {Herrnstein}, J., {Greenhill}, L., {Nakai}, N.,
  {Diamond}, P., \& {Inoue}, M. 1995, \nat, 373, 127

\bibitem[{{Neufeld} {et~al.}(1994){Neufeld}, {Maloney}, \&
  {Conger}}]{Neufeld1994}
{Neufeld}, D.~A., {Maloney}, P.~R., \& {Conger}, S. 1994, \apjl, 436, L127

\bibitem[{{Neufeld} \& {Melnick}(1991)}]{Neufeld1991}
{Neufeld}, D.~A. \& {Melnick}, G.~J. 1991, \apj, 368, 215

\bibitem[{{Peck} {et~al.}(2003){Peck}, {Henkel}, {Ulvestad}, {Brunthaler},
  {Falcke}, {Elitzur}, {Menten}, \& {Gallimore}}]{Peck2003}
{Peck}, A.~B., {Henkel}, C., {Ulvestad}, J.~S., {Brunthaler}, A., {Falcke}, H.,
  {Elitzur}, M., {Menten}, K.~M., \& {Gallimore}, J.~F. 2003, \apj, 590, 149

\bibitem[{{Phillips} {et~al.}(1980){Phillips}, {Kwan}, \&
  {Huggins}}]{Phillips1980}
{Phillips}, T.~G., {Kwan}, J., \& {Huggins}, P.~J. 1980, in IAU Symp. 87:
  Interstellar Molecules, 21--24


\bibitem[{{Prieto} {et~al.}(2004){Prieto}, {Meisenheimer}, {Marco}, {Reunanen},
  {Contini}, {Clenet}, {Davies}, {Gratadour}, {Henning}, {Klaas}, {Kotilainen},
  {Leinert}, {Lutz}, {Rouan}, \& {Thatte}}]{Prieto2004}
{Prieto}, M.~A., {et~al.}
  2004, \apj, 614, 135

\bibitem[{{Sawada-Satoh} {et~al.}(2000){Sawada-Satoh}, {Inoue}, {Shibata},
  {Kameno}, {Migenes}, {Nakai}, \& {Diamond}}]{Sawada2000}
{Sawada-Satoh}, S., {Inoue}, M., {Shibata}, K.~M., {Kameno}, S., {Migenes}, V.,
  {Nakai}, N., \& {Diamond}, P.~J. 2000, \pasj, 52, 421

\bibitem[{{Scoville}(1993)}]{Scoville1993}
{Scoville}, N. 1993, Mir Cookbook

\bibitem[{{Sofue}(1999)}]{Sofue1999}
{Sofue}, Y. 1999, Advances in Space Research, 23, 949



\bibitem[{{Trotter} {et~al.}(1998){Trotter}, {Greenhill}, {Moran}, {Reid},
  {Irwin}, \& {Lo}}]{Trotter1998}
{Trotter}, A.~S., {Greenhill}, L.~J., {Moran}, J.~M., {Reid}, M.~J., {Irwin},
  J.~A., \& {Lo}, K. 1998, \apj, 495, 740

\bibitem[{{Waters} {et~al.}(1980){Waters}, {Kakar}, {Kuiper}, {Roscoe},
  {Swanson}, {Rodriguez Kuiper}, {Kerr}, {Thaddeus}, \&
  {Gustincic}}]{Waters1980}
{Waters}, J.~W., {et~al.} 1980, \apj, 235, 57


\bibitem[{{Yamauchi} {et~al.}(2004){Yamauchi}, {Nakai}, {Sato}, \&
  {Diamond}}]{Yamauchi2004}
{Yamauchi}, A., {Nakai}, N., {Sato}, N., \& {Diamond}, P. 2004, \pasj, 56, 605

\bibitem[{{Yates} {et~al.}(1997){Yates}, {Field}, \& {Gray}}]{Yates1997}
{Yates}, J.~A., {Field}, D., \& {Gray}, M.~D. 1997, \mnras, 285, 303

\end{thebibliography}
\end{document}